\documentclass[preprint]{aastex}
\shorttitle{}
\shortauthors{}
\begin{document}

\title{Swift-XRT observation of 34 new INTEGRAL/IBIS AGNs: discovery of Compton thick 
and other peculiar sources.}

\author{A.~Malizia\altaffilmark{1},
R.~Landi\altaffilmark{1}, L.~Bassani\altaffilmark{1},A.~J.~Bird\altaffilmark{2}, M.~Molina\altaffilmark{2},
A.~De Rosa\altaffilmark{3}, M.~Fiocchi\altaffilmark{3},
N.~Gehrels\altaffilmark{4}, J.~Kennea\altaffilmark{5}, M.~Perri\altaffilmark{6}}

\altaffiltext{1}{IASF--Bologna/INAF, via P. Gobetti 101, 40129 Bologna, Italy}
\altaffiltext{2}{School of Physics and Astronomy, University of 
Southampton, Highfield, Southampton, SO17 1BJ, UK}
\altaffiltext{3}{IASF--Roma/INAF, via del Fosso del Cavaliere 100, 00133 
Roma, Italy}
\altaffiltext{4}{NASA/Goddard Space Flight Center, Greenbelt, MD 20771, USA}
\altaffiltext{5}{Department of Astronomy and Astrophysics, Pennsylvania State University, University Park, PA 16802, USA} 
\altaffiltext{6}{ASI Science Data Center, via G. Galilei, 00044 Frascati, Italy}

\begin{abstract}
For a significant number of the sources detected at high energies ($>$10 keV) by the INTEGRAL/IBIS and Swift/BAT instruments
there is either a lack information about them in the 2-10 keV range or they are totally unidentified. 
Herein, we report on a sample of 34 IBIS AGN 
or AGN candidate objects for which there is X-ray data in the Swift/XRT archive. 
Thanks to these X-ray follow up observations,
the identification of the gamma ray emitters has been possible and
the spectral shape in terms of photon index and absorption has been evaluated for the first time
for the majority of our sample sources.
The sample, enlarged to include 4 more AGN already discussed in the literature,
has been used to provide photon index and column density distribution.
We obtain a mean value of 1.88 with a dispersion of 0.12, i.e. typical of an AGN sample. 
Sixteen objects (47\%) have column densities in excess of 10$^{22}$
cm$^{-2}$ and, as expected, a large fraction of the absorbed sources
are within the Sey 2 sample.  We have provided a new diagnostic tool
(N$_{H}$ versus F$_{(2-10)keV}$/F$_{(20-100) keV}$ softness ratio) to
isolate peculiar objects; we find at least one absorbed Sey 1 galaxy, 3 Compton thick AGN candidates;
and  one secure example of a "true" type 2 AGN.
Within the sample of 10 still unidentified objects, 3 are almost
certainly AGN of type 2; 3 to 4 have spectral slopes typical of AGN;
and two are located high on the galactic plane and are strong enough radio emitters so that can 
be considered good AGN candidates.
\end{abstract}

\keywords{X-ray sources: general -- Galaxies: Seyfert}

\section{Introduction}
In the last few years, our knowledge of the hard X-ray sky ($>$10 keV)
has improved greatly thanks to the observations made by IBIS (Ubertini et al. 2003) 
on board INTEGRAL (Winkler et al. 2003) and
BAT (Barthelmy et al. 2005) on board Swift (Gehrels et al. 2004); both telescopes  operate in similar wavebands
(around 20-200 keV) 
with a sensitivity of about a  mCrab and a point source location accuracy of the order of a
few arcminutes.  These instruments have so far been used to survey the high energy sky in
a complementary way, as the first concentrates mainly on mapping the
galactic plane while the second mainly covers the high galactic latitude sky
so that together they will provide the best yet sample of objects
selected in the soft gamma-ray band.  
For a significant number of the sources detected by these two satellites there is either a
lack of information about them in the 2-10 keV range or they are totally unidentified.
For example, the third INTEGRAL/IBIS survey (Bird et al. 2007) contains 421 hard X-ray
(17-100 keV) emitters of which 113 are unidentified objects (26\% of the entire sample);
around 140 sources are AGN or AGN candidates and for almost half of these there is no information below 10 keV.
For all these objects, improved localization, as is now possible
with the current generation of focusing X-ray telescopes, is necessary in order
to pinpoint the optical counterpart and to definitely resolve their
nature/class.  Equally, spectral information in the
X-ray band is necessary to characterize the source spectral shape.
For these reasons, a program
of X-ray follow up observations of INTEGRAL/IBIS sources has recently
been initiated with the Swift satellite using the XRT instrument (0.2-10 keV, Burrows et al. 2006). 
Herein we report on a sample of 34 AGN or AGN candidates for which there is
X-ray data in the Swift/XRT archive up to the end of February 2007. 
For the majority of these sources the X-ray spectra are presented for the first time. 
Previously for a third of the sample only an estimate of the column densities have been reported by Sazonov et al. 2007; 
our results are in good agreement with theirs.
XRT follow up measurements allow in all cases an arcsecond positioning of the X-ray counterpart and in most
objects a definition of the X-ray spectral shape, in terms of photon index and
column density estimates. 
In particular, the study of the column density distribution in a sample of AGN selected above 10 keV is important so as
to quantify the fraction of objects missed by lower energy surveys, to find new Compton thick AGN or peculiar
sources and as an input parameter for synthesis models of the cosmic  X-ray background.  
All our sources have redshifts $z<$0.08 and so our results refer to the local Universe.

\section{Swift-XRT image analysis}
Due to the pointing strategy of Swift, short (a few ksec) repeated (up
to 5 times) measurements are typically performed for each target.  Table 1
lists for each INTEGRAL/IBIS source, the XRT position and the 90\% error box of the X-ray counterpart, 
the number of pointings$\footnote{In a few cases some observations have not been
  considered due to their poor quality and/or short duration ($<$ 1
  Ksec)}$, the total exposure time available, the mean count rates (i.e.
related to the sum of all available observations) of the detected
X-ray counterpart in the best energy range (see section 3) and, finally, 
its class and redshift (from NED unless otherwise stated).\\  
XRT data reduction was performed using the XRTDAS v. 2.0.1 standard data
pipeline package ({\sc xrtpipeline} v. 0.10.6), in order to produce
screened event files. All data are extracted only in the Photon
Counting (PC) mode (Hill et al. 2004), adopting the standard grade
filtering (0--12 for PC) according to the XRT nomenclature.  Images
have been extracted in the 0.3--10 keV band and searched for
significant excesses falling within the INTEGRAL/IBIS 90\% confidence
circle as reported in Bird et al. (2007); in all cases a single X-ray
source was detected either inside this uncertainty circle or just at its
border.  All sources are detected above 3$\sigma$ significance level
and in most cases are quite bright at X-ray energies, so that even a
short exposure can provide useful information. \\ 
The overall sample consists of 11 Seyferts of type 2 (Sey 2), 13 Seyferts of type 1 (Sey 1)
(i.e. type in the range 1-1.5) including two Narrow Line
Seyfert 1s.\\
10 sources are still optically unclassified and their
extragalactic nature is inferred by indirect arguments.  3 objects
(IGR J14492-5535, IGR J15539-6142 and IGR J20186+4043) are associated
with galaxies (2MASX J14491283-5536194, ESO 136-G006 and 2MASX J20183871+4041003 respectively) 
in which the nuclear activity was unknown before X/gamma-ray measurements; 
their AGN class is still uncertain but the presence of strong absorption in the X-ray spectra
implies a type 2 nature
(see also section 3).  Three other objects (SWIFT J0216.3+5128, IGR
J13109-5552 and SWIFT J16563-3302) are located 6 degrees above the
galactic plane and are all radio emitting objects besides being strong
high energy emitters; in particular IGR J13109-5552 and SWIFT J16563-3302,
associated with PMN J1310-5552 and NVSS J165616-330211 respectively,
are quite strong at radio frequencies (353 mJy at 4.85 MHz for the former
and 410 mJy at 20 cm for the latter). Historically AGN were discovered by
radio observations, i.e. radio selection is often a way in which to
recognize active galaxies, except at lower luminosities where also
starburst galaxies emit at radio frequencies. Therefore for bright
objects located away from the galactic plane, mere detection in radio
provides support for the presence of an active nucleus.  Contamination
from galactic sources may come from pulsars, microquasars and
supernova remnants (Molina 2004) so that it is less easy to claim an
AGN association for radio objects on the galactic plane such as IGR
J14003-6362, IGR J18259-0706, IGR J19443+2117 and IGR J21178+5139. 
However, IGR J19443+2117 is a flat spectrum radio
source according to NED while IGR J21178+5139 is also a 2MASS extended
object although not recognized as a galaxy (typically the most likely
association in this catalogue).  Clearly all these sources must await
optical (or even infrared) spectroscopy in order to be definitely confirmed as
AGN.
In the meantime X-ray spectroscopy can shed light on their extragalactic nature.\\

\section{Swift-XRT spectral analysis}
Events for spectral analysis were extracted within a circular region of radius 20$^{\prime \prime}$, 
which encloses about 90\% of the PSF at 1.5 keV (Moretti et al. 2004) 
centered on the source position.
The background was extracted from various source-free regions close to the 
X-ray source of interest using both circular/annular regions of various
radii, in order to ensure an evenly sampled background. In all 
cases, the spectra were extracted from the corresponding event files using 
{\sc XSELECT} software and binned using {\sc grppha} so that the $\chi^{2}$ 
statistic could reliably be used. We used the 
latest version (v.008) of the response matrices and created individual 
ancillary response files (ARF) using {\sc xrtmkarf}. Spectral analyses 
have been performed using XSPEC version 12.2.1.
For those sources with more than one pointing, we performed the spectral 
analysis of each observation  and then of  the combined spectra, in order to improve the statistical 
quality of the data; in most cases spectra from the individual pointings were compatible with each other within the
respective uncertainties, thus justifying a combined analysis.
Only in 5 cases (see sources labeled with 'v' in table 2), do we find evidence for 
variability between measurements but always in flux 
and not in spectral shape; clearly for these five sources the reported 2-10 keV flux is an average
over a number of observations. 
Due to the low statistics available, we have identified for each source the best energy range 
for the spectral analysis  and
have employed a simple power law absorbed by both a Galactic 
and an intrinsic column density as our baseline model. 
The results of the analysis of the combined observations  are
reported in table 2 where we list the energy band used for the 
spectral fit, the Galactic absorption according to Dickey \& Lockman 
(1990), the photon index, the column density in excess to the Galactic value, if any, 
the 2--10 keV flux and the reduced $\chi^{2}$ of the fit.
In the last column of table 2 we also list  the IBIS 20--100 keV flux as reported by Bird et al.(2007).
In a few cases the photon index was fixed to the
canonical AGN value ($\Gamma=1.8$) so as to allow a measurement of the intrinsic column density; 
in the particular case of IGR J14175-4641 
only a rough estimate of the 2-10 keV flux was allowed by the data.
The presence of the iron line around 6.4 keV, a typical feature of an AGN spectrum,
is not statistically required in any of the spectra analyzed but  this is definitely due to the instrument 
sensitivity and lack of good coverage at these energies.
For those sources where there is coverage at low energies ($<$1 keV) we have checked for the presence of possible soft excess component. 
In only one case, the Seyfert 1 galaxy ESO 323-G077, do we find significant 
evidence for this extra component which could be equally well fitted with an unabsorbed power law having the same photon index of the primary continuum
or a black body model with kT $\sim$ 0.2 keV (see table 2).
All quoted errors correspond to  90$\%$ confidence level for a single 
parameter of interest ($\Delta\chi^{2}=2.71$).

\section{Photon index and column density distributions}
Despite the short exposures available,
it is still possible to obtain information on the spectral shape of the objects analyzed.  In
particular, we concentrate here on the photon index distribution and
on the absorption properties.  To enlarge the
sample, we have included four more XRT/IBIS AGNs: SWIFT J0601.9-8636
(Sey 2, z=0.006), SWIFT J1009.3-4250 (Sey 2, z=0.033), SWIFT
J1038.8-4942 (Sey 1.5, z=0.060) and SWIFT J1238.9-2720 (Sey 2,
z=0.024), already discussed by Landi et al. (2007).  Their spectral
data have been re-analyzed in the same way as for the other sources
presented in
this work and the results are included in the following discussion.\\
The distribution of photon indices for those AGN with optical
classification is shown in figure 1; sources where $\Gamma$ was not
constrained by the data were not considered.  It is well known that
the distribution of X-ray spectral slopes of AGN peaks around 1.9 and
has a non-negligible dispersion (0.2-0.3, see for example Mateos et
al. 2005 and references therein). We obtain 
a mean $\Gamma$ value of 1.88  with a dispersion  
of 0.12, i.e. most of our objects have spectra close to the canonical AGN value.
Flat ($\Gamma<$1.5) spectra are far less common, especially in
view of the large uncertainties associated with the XRT spectra;
nevertheless they exist for some sources, such as IGR J21247+5058 as has been
confirmed by a recent XMM observation (Molina et al. in preparation).
The photon index distribution can also be used to characterize unclassified objects: 
SWIFT J0216.3+5128, IGR J14003-6326 and IGR J19443+2117 and
possibly IGR J21178+5139 have spectral slopes typical of AGN
supporting their extragalactic nature. Others have flat spectra
(1.1-1.5) which are unusual, but still compatible with an AGN association.

\begin{figure}
\plotone{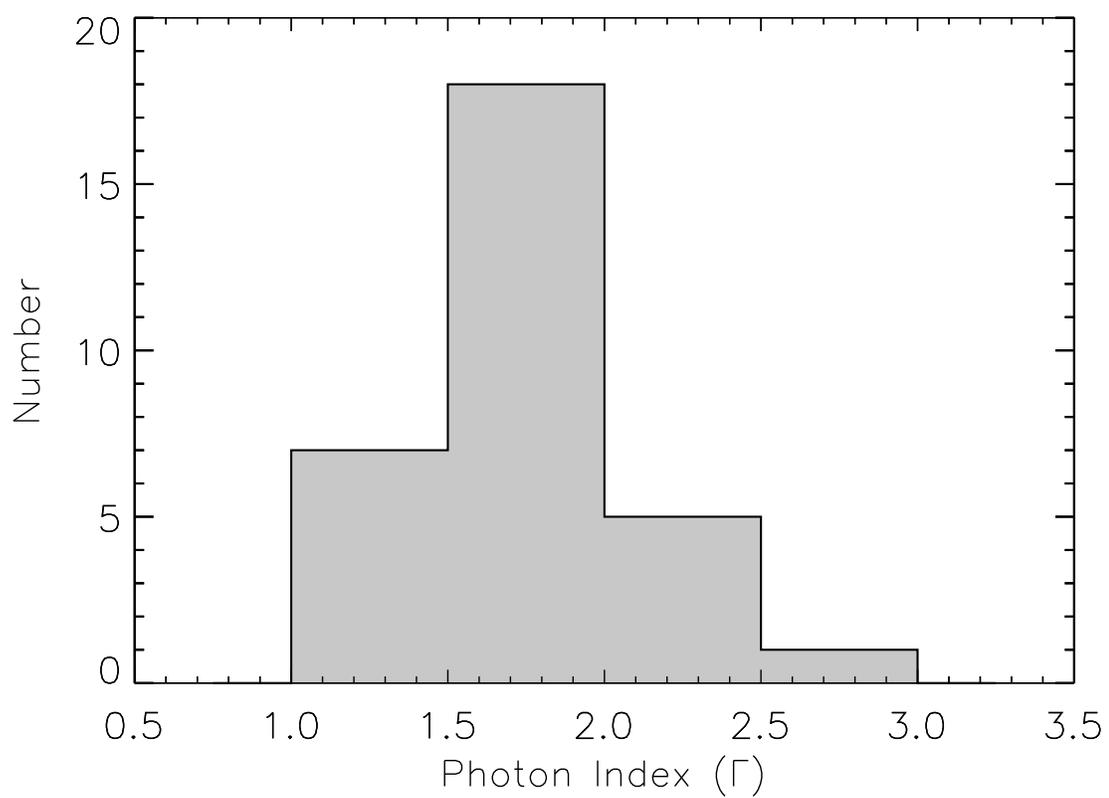}
\caption{Distribution of the photon index of the AGN with
optical identifications.}
\label{Fig1}
\end{figure}

The column density distribution for each class of objects (type 1, type
2 and unclassified), are shown in figures 2, 3 and 4 respectively.  As
we can see from table 2, various sources of the sample (8 Sey 1, 3 Sey
2 and 1 unclassified) do not show intrinsic absorption and therefore
their Galactic column density has been taken as an upper limit to the
value of N$_{H}$; these sources are reported in the histograms as
shaded areas.  Absorption in excess of the galactic value has been
measured in 25 objects (around 66\% of the sample), but only 16 (47\%)
have a column density in excess of 10$^{22}$ cm$^{-2}$ (typically
taken as the dividing line between absorbed and unabsorbed objects
$\footnote{This N$_{H}$ value represents the amount of absorbing neutral
  gas needed to hide the broad emission-line region (BLR), assuming a
  Milky Way gas-to-dust ratio.}$.  Absorption was found in 15\% of
type 1 AGN, in 73\% of type 2 AGN and in 60\% of unclassified objects.
Indeed the column density distribution of type 1 and type 2 AGN is
significantly different: while type 1 objects cluster in the
10$^{21}$-- 10$^{22}$ cm$^{-2}$ range, type 2 sources have a much wider
distribution. The distribution of unclassified objects resembles that
of Sey 2s suggesting that many of these sources could be of this type.

\begin{figure}
\plotone{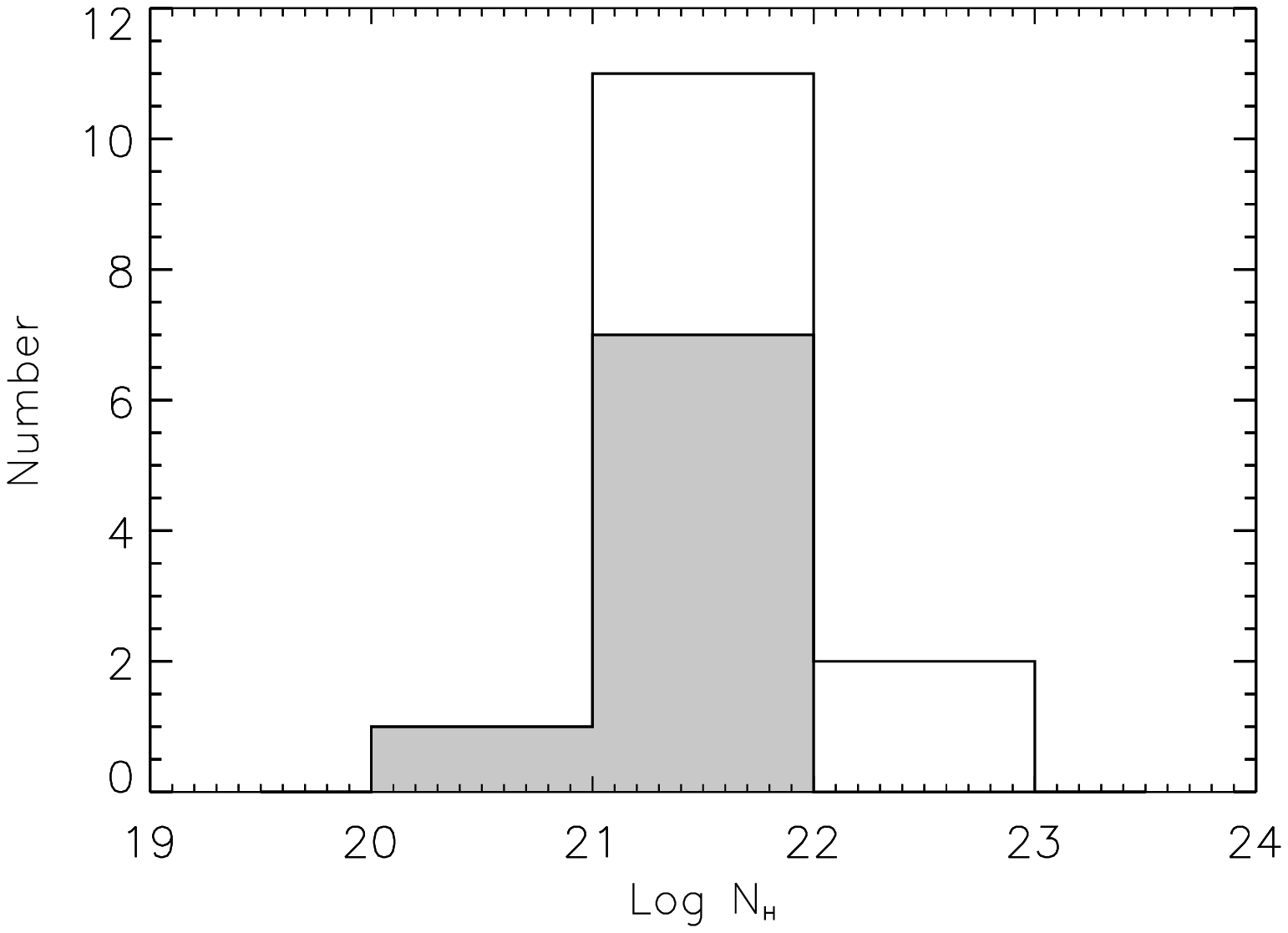}
\caption{Column density distribution of the type 1 Seyferts of our sample, shaded areas 
are for upper limits.}
\label{Fig1}
\end{figure}

\begin{figure}
\plotone{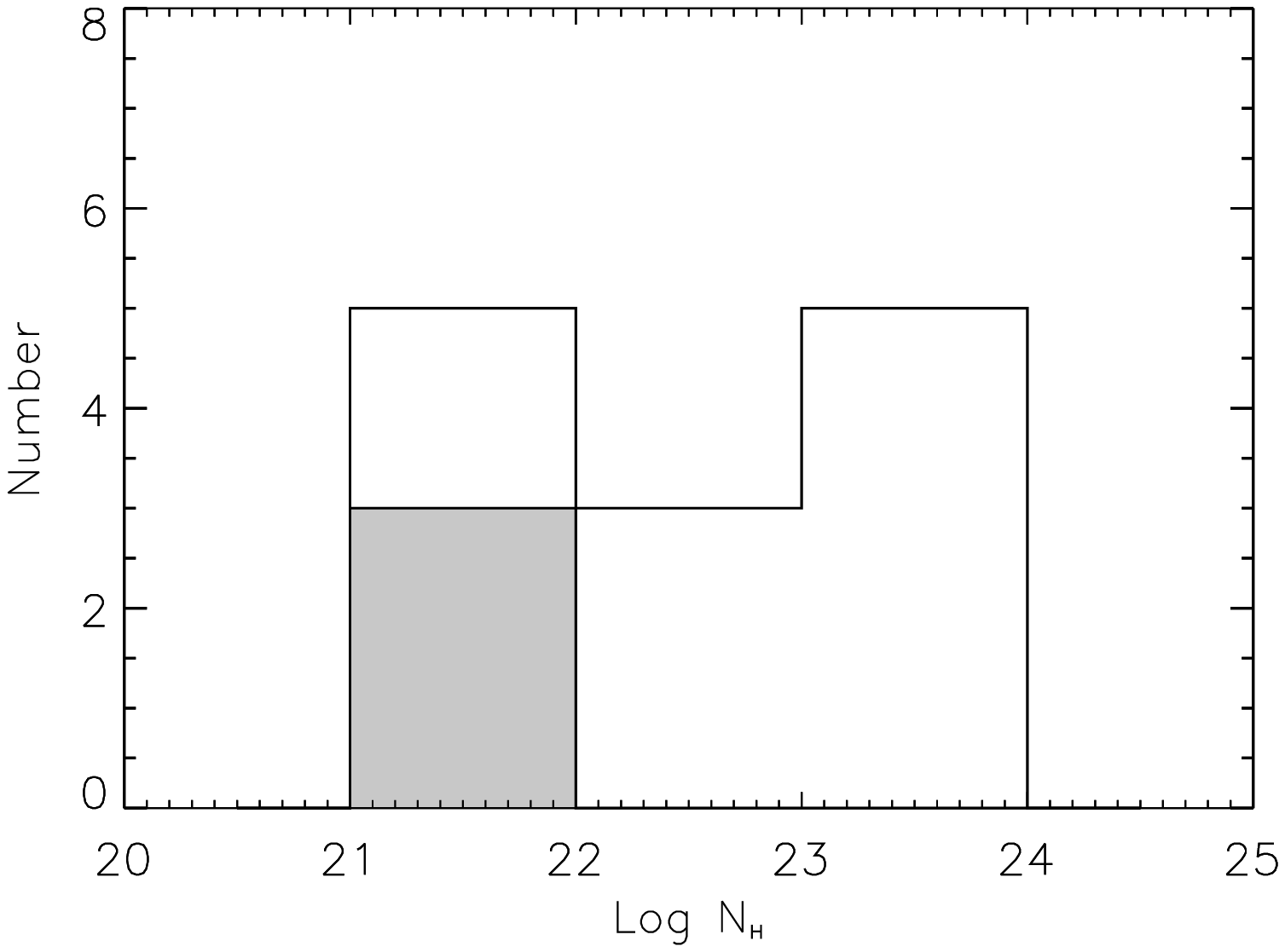}
\caption{Column density distribution of the type 2 Seyferts of our sample, shaded areas 
are for upper limits.}
\label{Fig1}
\end{figure}

\begin{figure}
\plotone{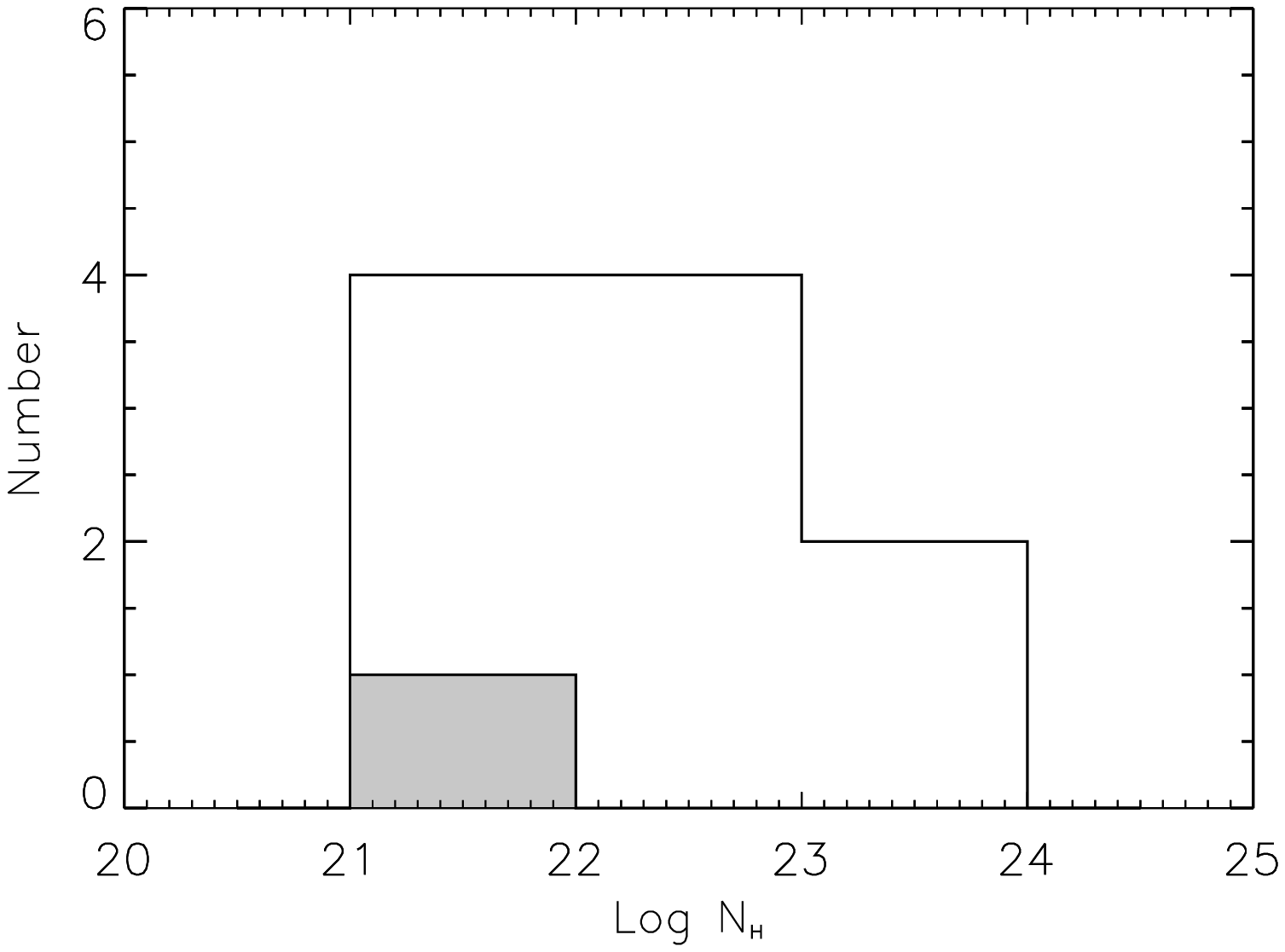}
\caption{Column density distribution of the still unclassified subset of objects, shaded areas 
are for upper limits.}
\label{Fig1}
\end{figure}

\section{Discussion}
The basic hypothesis of the unified theory of AGN is that the X-ray
absorption and optical obscuration are heavily related: absorbed AGN
should be classified as type 2 in optical (with the Broad Line Region
hidden) while unabsorbed ones are expected to be of type 1 (with the
Broad Line Region visible). This relationship is not always
respected as is evident in our sample where we find absorption in
two Sey 1s (ESO 323-G077 and IGR J21247+5058) and lack of it in a
number of Sey 2s.  While IGR J21247+5058 is located on the Galactic
plane so that the excess absorption may be due to extra material along
the line of sight, ESO 323-G077 is high in galactic latitude and so
likely to be intrinsically absorbed.  
Evidence of heavly absorption was first reported by Sazonov and Revnivtsev (2004)
in the RXTE survey, also this AGN is peculiar for
being one of the rare cases of a high scattering polarization Sey 1
(Schmidt et al. 2004).  Within the standard model of Seyfert nuclei,
both peculiarities of ESO323-G077 can be well understood by assuming that
this AGN is observed at an inclination angle where the nucleus is
partially obscured by the torus (hence the high absorption measured in
X-rays) and is seen mainly indirectly in the light scattered by dust
clouds within or above the torus (hence the high linear polarization
observed). Further X-ray observations of this object are clearly
desirable because of the constraints that can be derived on the
geometric properties of the absorbing region.  Contrary to unified
models, we have also found 5 Sey 2 with no clear evidence of
absorption in their X-ray spectra.  Several authors have reported
similar examples of AGN with no broad emission lines and low N$_{H}$
(Panessa and Bassani 2002, Corral et al 2005).  
There are various explanations for this discrepancy including the low statistical
quality of the X-ray spectra and the non simultaneity of the optical
and X-ray observations.  
The low X-ray exposures available are clearly the most important
source of uncertainty. 
Alternatively, it is possible that these Sey 2 are either Compton thick type 2 AGN or
"true" Sey 2, i.e. objects where the Broad Line Region is not hidden but
likely not to be present. 
If the source is Compton thick, the 2--10 keV emission is completely blocked and could
be seen as reflected/scattered radiation with no apparent absorption.
On the other hand the soft gamma-ray photons are unaffected by
absorption as long as logN$_{H}$ is below 24.5; for higher column
densities, even emission above 20 keV is blocked and can only be seen
indirectly.  Generally indirect arguments allow discrimination between the
Compton thick and Compton thin nature of type 2 AGN, such as the
equivalent width of the iron line and the ratio of isotropic versus
anisotropic luminosities typically L$_{X}$/L$_{[OIII]}$ $\footnote{where the
  L$_{[OIII]}$ is corrected for reddening in the host galaxy by means
  of the H$_{\alpha}$/H$_{\beta}$ Balmer decrement}$ (Bassani et al. 1999
and Panessa and Bassani 2002). Unfortunately, we do not have such
information for most of our sources and so we have to rely on a new
diagnostic diagram.  Figure 5 is a plot of the absorption versus the
flux softness ratio (F$_{(2-10) keV}$/F$_{(20-100) keV}$): a clear trend
of decreasing softness ratio as the absorption increases is visible as
expected if the 2-10 keV flux is progressively depressed as the
absorption becomes stronger.  Open symbols in the figure represent
upper limits on the column density, while lines represent the expected
values for an absorbed power law with photon index of 1.5 and
1.9.  As anticipated, most of our sources follow the expected
trend except for a few objects which have a too low softness ratio for
the observed column density, suggesting a Compton thick nature.
Indeed if we take the Swift/XRT observation of MKN 3, a confirmed Compton
thick Sey 2, and treat it as one of our sources,
we find that it is located in the same region, while the true column density 
of this source is 1.3 $\times$ 10$^{24}$ (Cappi et al. 1999).\\
We therefore suggest that IGR J14175-4641, IGR J16351-5806 and SWIFT
J0601.9-8636 are new Compton thick AGN candidates where further, more
sensitive X-ray observations, particularly around the iron line could
help in confirming these results.  Similarly IGR J07565-4139 and IGR
J18259-0706 are outside the expected trend, although their softness
ratio is not so dramatically low and could be explained by the
uncertainties associated with the flux estimates as well as by
variability in the source since the X-ray and soft gamma-ray fluxes
are not taken simultaneously.  While for IGR J18259-0706 we must await
the source classification to confirm its extragalactic nature and
therefore the presence of any peculiarity, IGR J07565-4139 is
intriguing as the soft X-ray emission is unlike any other seen in a small
sample of absorbed AGN (De Rosa et al. in preparation). 
Two more unabsorbed Sey 2, IGR J12415-5750 and IGR J14515-5542, are also
anomalous but are unlikely to be Compton thick objects as their location
is well within the expected region of N$_{H}$ versus F$_{(2-10) keV}$/F$_{(20-100) keV}$; 
while IGR J12415-5750 may have been
misclassified in optical (see http://www.astro.umd.edu/$\sim$lwinter/research/BAT/xray.html),  
IGR J14515-5542 shows a true Sey 2 spectrum and furthermore has a
L$_{X}$/L$_{[OIII]}$ ratio typical of unabsorbed type 1 sources
(Masetti et al. 2006) and so qualifies as a "true" Sey 2 nucleus;
indeed the observed column density barely explains the occultation of
its Broad Line Region.
 
\section{Conclusions}
Swift-XRT follow-up X-ray observations of 34 INTEGRAL/IBIS AGN and AGN
candidates have pinpointed their X-ray counterparts allowing the
unambiguous identification of the soft gamma-ray source and the
determination of the AGN class in most cases.  Furthermore, for most
objects in the sample, the photon index and absorption have been measured.
When this sample is enlarged to include 4 more AGN already discussed in the literature,
we find a mean photon index of 1.88 with a dispersion of 0.12.
47\% of the objects have column densities in excess of 10$^{22}$
cm$^{-2}$ and, as expected, a large fraction of the absorbed sources
are within the Sey 2 sample.  We have provided a new diagnostic tool
(N$_{H}$ versus F$_{(2-10) keV}$/F$_{(20-100) keV}$ softness ratio) with which to
isolate or find a few peculiar objects: a) 2 absorbed Sey 1 galaxies (ESO
323-G077 and IGR J21247+5058); b) 3 Compton thick AGN candidates
(SWIFT J0601.98636, IGR J14175-4841 and IGR J16351-5806); and c) at
least one secure example (IGR J14415-5542) of a "true" type 2 AGN,
i.e. one where the Broad Line Region is not hidden but likely not to be
present.  Within the sample of 10 still unidentified objects, 3 (IGR
J14492-5535, IGR J15539-6142 and IGR J20186+4043) are almost
certainly AGN of type 2; 3 to 4 more (SWIFT J0216.3+5128, IGR
J14003-6326 and IGR J19443+2117 and possibly IGR J21178+5139) have
spectral slopes typical of AGN (with the first two likely to be of
type 2 because of the measured absorption).  Two objects (IGR
J13109-5552 and SWIFT J16563-3302) are located high enough off the
galactic plane and are strong radio emitters and are therefore good AGN
candidates; only IGR J18259-0706 has a doubtful AGN association
despite being a radio source.

\acknowledgments 
We acknowledge the Italian Space Agency financial and programmatic support via
contracts I/R/046/04 and I/023/05/0.\\
A.M. thanks J. B. Stephen for a careful reading of the manuscript.

\begin{figure}
\plotone{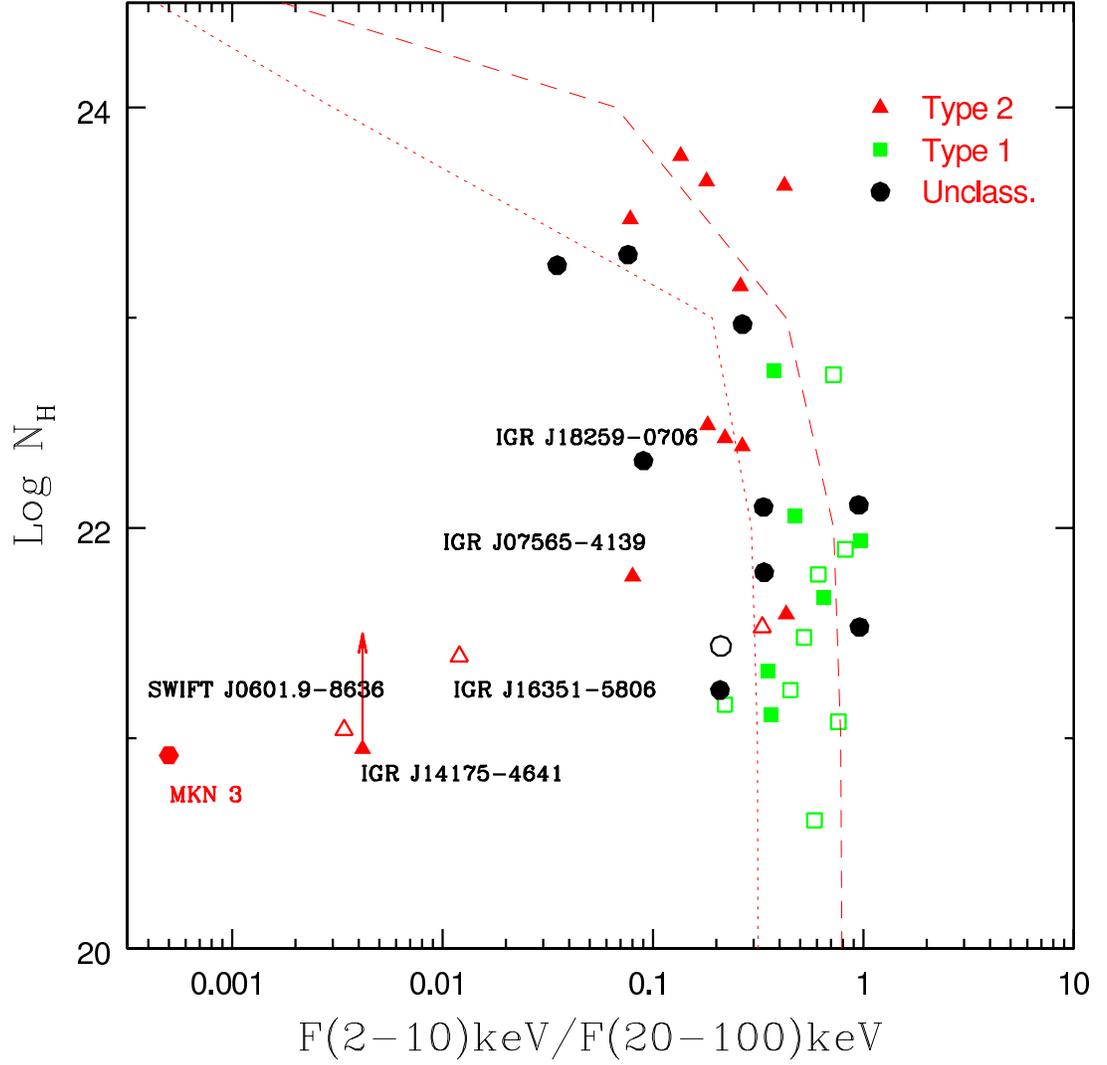}
\caption{F$_{2-10 keV}$/ F$_{20-100 keV}$ flux ratio of our enlarged sample. Open symbols
represent upper limit values of the column densities and lines correspond to expected values for an absorbed power law
with photon index 1.5 (dot) and 1.9 (dash).}
\label{Fig2}
\end{figure}

\clearpage

\begin{table}
\begin{center}
\footnotesize
\caption{\emph{Swift}/XRT positions}
\begin{tabular}{lccccrcll}
\tableline
\tableline

 Source            &      RA     &    Dec        & error & Obs.   & Expo.&    count rate      & class & redshift  \\     
                          &                &                    & $\it{arcsec}$&    & sec    &                 cts/s     &       \\ 
\tableline                                                                                                  
IGR J02097+5222    & 02 09 37.68 & +52 26 43.56  & 3.54 &  2 & 8034  & 0.366$\pm$0.0067  & Sey 1        & 0.049    \\
SWIFT J0216.3+5128 & 02 16 27.01 & +51 25 25.50  & 3.56 &  3 & 16401 & 0.169$\pm$0.0032  & ??           & \nodata  \\                     
NGC 1142           & 02 55 12.19 & -00 11 02.31  & 3.80 &  4 & 18807 & 0.034$\pm$0.0013  & Sey 2        & 0.029    \\
1H 0323+342        & 03 24 41.04 & +34 10 45.49  & 3.53 &  3 & 19510 & 0.368$\pm$0.0043  & Sey 1        & 0.061    \\
LEDA 168563        & 04 52 04.85 & +49 32 43.72  & 3.59 &  1 & 1111  & 0.297$\pm$0.0016  & Sey 1        & 0.029    \\
IGR J07565-4139    & 07 56 19.71 & -41 37 41.62  & 3.75 &  1 & 14750 & 0.020$\pm$0.0012  & Sey 2$^{a}$  & 0.021    \\
IGR J07597-3842    & 07 59 41.66 & -38 43 57.35  & 3.52 &  4 & 19753 & 0.383$\pm$0.0044  & Sey 1.2$^{a}$& 0.040    \\
SWIFT J0917.2-6221 & 09 16 09.01 & -62 19 29.04  & 3.56 &  2 & 12893 & 0.202$\pm$0.0039  & Sey 1        & 0.057    \\
IGR J10404-4625    & 10 40 22.27 & -46 25 24.69  & 3.61 &  4 & 19561 & 0.064$\pm$0.0018  & Sey 2$^{b}$  & 0.024    \\
IGR J12391-1612    & 12 39 05.98 & -16 10 48.14  & 3.62 &  1 & 6625  & 0.106$\pm$0.0046  & Sey 2        & 0.037    \\
IGR J12415-5750    & 12 41 25.36 & -57 50 03.97  & 3.57 &  2 & 12843 & 0.137$\pm$0.0033  & Sey 2        & 0.024    \\
ESO 323-G077       & 13 06 26.13 & -40 24 53.22  & 3.72 &  2 & 7953  & 0.018$\pm$0.0015  & Sey 1.2      & 0.015    \\
IGR J13109-5552    & 13 10 43.08 & -55 52 11.66  & 3.73 &  2 & 6563  & 0.092$\pm$0.0037  &  ??          & \nodata  \\
IGR J14003-6326    & 14 00 45.35 & -63 25 41.17  & 3.77 &  2 & 5010  & 0.174$\pm$0.0071  & ??           & \nodata  \\
IGR J14175-4641    & 14 17 03.94 & -46 41 39.06  & 5.70 &  1 & 5541  & 0.009$\pm$0.0007  & Sey 2$^{a}$  & 0.076    \\
IGR J14471-6319    & 14 47 14.69 & -63 17 19.56  & 4.03 &  1 & 3911  & 0.033$\pm$0.0029  & Sey 2$^{c}$  & 0.038    \\
IGR J14492-5535    & 14 49 12.67 & -55 36 20.32  & 3.77 &  1 & 3606  & 0.016$\pm$0.0022  & ??           & \nodata  \\
IGR J14515-5542    & 14 51 33.43 & -55 40 39.41  & 3.60 &  3 & 12515 & 0.096$\pm$0.0028  & Sey 2$^{a}$  & 0.018    \\
IGR J14552-5133    & 14 55 17.36 & -51 34 15.68  & 3.61 &  2 & 8423  & 0.216$\pm$0.0051  & NLSey1$^{a}$ & 0.016    \\
IGR J15539-6142    & 15 53 35.22 & -61 40 55.40  & 4.71 &  1 & 4840  & 0.006$\pm$0.0013  &  ??          & 0.015    \\
IGR J16351-5806    & 16 35 13.42 & -58 04 49.69  & 5.08 &  2 & 6233  & 0.006$\pm$0.0009  & Sey 2        & 0.009    \\
IGR J16482-3036    & 16 48 14.94 & -30 35 06.06  & 3.54 &  1 & 9205  & 0.209$\pm$0.0048  & Sey 1        & 0.031    \\
IGR J16558-5203    & 16 56 05.73 & -52 03 41.18  & 3.52 &  1 & 12850 & 0.438$\pm$0.0060  & Sey 1.2$^{a}$& 0.054    \\
SWIFT J1656.3-3302 & 16 56 16.83 & -33 02 12.18  & 3.73 &  2 & 9152  & 0.071$\pm$0.0028  &  ??          & \nodata  \\
IGR J17488-3253    & 17 48 54.82 & -32 54 47.83  & 3.53 &  1 & 16780 & 0.196$\pm$0.0035  & Sey 1$^{a}$  & 0.020    \\
IGR J18244-5622    & 18 24 19.49 & -56 22 08.72  & 4.39 &  1 & 1631  & 0.047$\pm$0.0054  & Sey 2$^{a}$  & 0.017    \\
IGR J18259-0706    & 18 25 57.25 & -07 10 24.54  & 3.71 &  1 & 4260  & 0.055$\pm$0.0036  &  ??          & \nodata  \\
IGR J19443+2117    & 19 43 56.20 & +21 18 22.95  & 3.53 &  1 & 11080 & 0.267$\pm$0.0049  &  ??          & \nodata  \\
IGR J20186+4043    & 20 18 38.46 & +40 40 59.16  & 4.81 &  1 & 2439  & 0.008$\pm$0.0019  &  ??          & \nodata  \\
IGR J20286+2544    & 20 28 35.05 & +25 44 01.57  & 4.61 &  1 & 45385 & 0.012$\pm$0.0016  & Sey 2$^{d}$  & 0.014    \\
IGR J21178+5139    & 21 17 47.24 & +51 38 53.62  & 10.0 &  2 & 5362  & 0.016$\pm$0.0018  &  ??          & \nodata  \\
IGR J21247+5058    & 21 24 39.44 & +50 58 24.41  & 3.52 &  2 & 10190 & 0.428$\pm$0.0065  & Sey 1        & 0.020    \\
SWIFT J2127.4+5654 & 21 27 45.58 & +56 56 35.64  & 3.53 &  2 & 10130 & 0.374$\pm$0.0061  & NLSey1$^{e}$  & 0.014    \\
NGC 7603           & 23 18 56.69 & +00 14 38.22  & 3.51 &  3 & 22860 & 0.494$\pm$0.0047  & Sey 1.5      & 0.029    \\                      
\tableline                                                                                                   
\end{tabular}                                                                                                  
Notes: a) Masetti et al. 2007; b) Masetti et al. 2006; c) Masetti et al. 2006b; d) Masetti et al. 2006a; 
e) Halpern Atel 847.
\end{center}
\end{table}

\begin{table}
\begin{center}
\footnotesize
\caption{\emph{Swift}/XRT spectra}
\begin{tabular}{lccccclc}
\tableline
\tableline
Source   & band  &    N$_{HGal}^{\dagger}$            &  $\Gamma$   &  N$_{H}^{\dagger}$         & F$_{(2-10) keV}^{\ddagger}$    & $\chi^{2}$/d.o.f   & F$_{(20-100) keV}^{\ddagger}$  \\ 
         &  keV  &    $\times$ 10$^{21}$   &             & $\times$ 10$^{22}$ & $\times$ 10$^{-11}$         &                    &  $\times$ 10$^{-11}$   \\
\hline
\hline
IGR J02097+5222              & 0.5-7.0  & 1.69 & 1.88$^{+0.06}_{-0.06}$  & \nodata                  & 1.30 & 107.5/129   & 2.90 \\
SWIFT J0216.3+5128           & 1.0-8.0  & 1.58 & 1.91$^{+0.13}_{-0.13}$  & 1.27$^{+0.16}_{-0.14}$   & 1.27 & 106.9/117 & 3.79 \\
NGC 1142                     & 1.0-7.0  & 0.64 & 1.8 (fixed)  & 44.9$^{+4.3}_{-4.0}$                & 0.96 & 33.7/26   & 5.33 \\
1H 0323+342                  & 0.2-7.0  & 1.44 & 2.13$^{+0.03}_{-0.04}$  & \nodata                  & 0.94 & 179.7/185 & 4.21 \\
LEDA 168563                  & 0.8-7.0  & 5.42 & 2.06$^{+0.18}_{-0.17}$  & \nodata                  & 4.54 & 10.2/16   & 6.31 \\
IGR J07565-4139              & 1.0-6.0  & 4.73 & 1.73$^{+0.43}_{-0.47}$  & 0.59$^{+0.45}_{-0.42}$   & 0.15 & 27.6/27   & 1.78 \\
IGR J07597-3842$^{v}$        & 1.0-7.2  & 6.04 & 1.80$^{+0.04}_{-0.04}$  &  \nodata                 & 2.37 & 241.5/259 & 3.89 \\
SWIFT J0917.2-6221           & 1.0-7.0  & 1.91 & 1.57$^{+0.16}_{-0.14}$  & 0.47$^{+0.15}_{-0.13}$   & 1.44 & 180.1/121 & 2.20 \\
IGR J10404-4625$^{v}$        & 1.0-7.0  & 1.36 & 1.47$^{+0.30}_{-0.27}$  & 2.67$^{+0.65}_{-0.54}$   & 0.72 & 90.0/80   & 3.27 \\
IGR J12391-1612              & 1.0-8.5  & 0.37 & 1.72$^{+0.36}_{-0.32}$  & 3.08$^{+0.84}_{-0.67}$   & 1.11 & 33.7/43   & 6.26 \\
IGR J12415-5750              & 1.0-6.0  & 3.45 & 1.63$^{+0.08}_{-0.07}$  &  \nodata                 & 0.77 & 69.6/79   & 2.32 \\
ESO 323-G077$^{v,\blacktriangle}$   & 0.5-6.0  & 0.70 & 2.40$^{+0.44}_{-0.44}$  & 6.61$^{+1.24}_{-1.18}$   & 1.03 & 41.5/50   & 2.94 \\
IGR J13109-5552              & 0.5-7.0  & 2.76 & 1.50$^{+0.12}_{-0.12}$ &  \nodata                  & 0.51 & 40.9/54   & 2.46 \\
IGR J14003-6326              & 1.0-6.0  & 15.0 & 2.21$^{+0.39}_{-0.38}$ & 1.28$^{+0.64}_{-0.62}$    & 1.43 & 34.3/40   & 1.51 \\     
IGR J14175-4641$^{\star}$    & \nodata  & \nodata & \nodata & \nodata               & 0.006 & \nodata & 1.41 \\
IGR J14471-6319              & 1.0-7.0  & 58.0 & 1.70$^{+1.04}_{-1.03}$  & 2.44$^{+2.06}_{-1.76}$   & 0.39 & 5.8/10   & 1.47 \\ 
IGR J14492-5535              & 2.0-8.5  & 4.99 & 1.8 (fixed) &  9.3$^{+6.0}_{-4.8}$                 & 0.42 & 19.0/24   & 1.58 \\
IGR J14515-5542              & 1.0-6.0  & 5.31 & 1.37$^{+0.20}_{-0.20}$  & 0.39$^{+0.18}_{-0.16}$   & 0.71 & 42.0/54  & 1.65 \\
IGR J14552-5133              & 0.5-6.0  & 3.37 & 1.93$^{+0.07}_{-0.07}$  & \nodata                  & 0.89 & 73.1/81   & 1.17 \\
IGR J15539-6142              & 0.7-7.0  & 3.01 & 1.8 (fixed) & 17.6$^{+32.8}_{-13.0}$               & 0.073& 10.6/11   & 2.13 \\
IGR J16351-5806              & 1.0-6.0  & 2.47 & 1.61$^{+0.75}_{-0.68}$  & \nodata                  & 0.031& 4.4/5     & 1.98 \\
IGR J16482-3036              & 0.5-8.3  & 1.76 & 1.71$^{+0.11}_{-0.12}$  & 0.13$^{+0.5}_{-0.06}$    & 1.13 & 80.8/96   & 3.12 \\
IGR J16558-5203              & 0.4-8.6  & 3.04 & 1.85$^{+0.06}_{-0.04}$  & \nodata                  & 1.77 & 221.0/217 & 3.39 \\
SWIFT J1656.3-3302           & 0.7-7.0  & 2.22 & 1.36$^{+0.23}_{-0.22}$  & 0.17$^{+0.15}_{-0.14}$   & 0.49 & 50.5/61   & 2.36 \\
IGR J17488-3253              & 0.5-7.7  & 5.30 & 1.60$^{+0.12}_{-0.10}$  & 0.21$^{+0.08}_{-0.07}$   & 1.40 & 133.1/137 & 4.03 \\
IGR J18244-5622              & 2.0-8.0  & 0.75 & 1.94$^{+1.86}_{-1.39}$  & 14.1$^{+15.4}_{-7.9}$    & 0.67 & 12.0/13   & 2.54 \\
IGR J18259-0706              & 1.0-7.0  & 7.10 & 1.40$^{+0.51}_{-0.47}$  & 0.62$^{+0.40}_{-0.33}$   & 0.54 & 13.2/21   & 1.60 \\
IGR J19443+2117              & 0.5-8.2  & 8.28 & 1.96$^{+0.12}_{-0.11}$  & 0.34$^{+0.13}_{-0.11}$   & 1.85 & 139.0/127 & 1.93 \\
IGR J20186+4043              & 1.0-8.0  & 12.0 & 1.8 (fixed) & 20.0$^{+20.2}_{-12.3}$               & 0.17 & 2.8/6     & 2.24 \\
IGR J20286+2544              & 1.0-7.0  & 2.62 & 1.8 (fixed) & 42.3$^{+20.8}_{-19.5}$               & 2.35 & 12.4/14   & 5.64 \\
IGR J21178+5139              & 1.0-6.2  & 14.3 & 1.8 (fixed) & 2.11$^{+1.52}_{-1.03}$               & 0.21 & 21.4/27   & 2.26 \\
IGR J21247+5058$^{v}$        & 1.0-8.5  & 11.1 & 1.44$^{+0.12}_{-0.12}$ & 1.15$^{+0.27}_{-0.25}$    & 5.20 & 196.2/175 & 11.02 \\
SWIFT J2127.4+5654           & 0.8-7.0  & 7.87 & 1.88$^{+0.05}_{-0.05}$ &  \nodata                  & 2.20 & 187.2/181 & 2.71 \\
NGC 7603$^{v}$               & 0.2-7.5  & 0.41 & 2.05$^{+0.03}_{-0.04}$ &  \nodata                  & 2.40 & 375.4/231 & 4.10 \\
\tableline
SWIFT J0601.9-8636           & 0.5-5.0  & 1.10 & 1.08$^{+2.30}_{-1.08}$ & \nodata                  & 0.011 & 3.5/2       &   3.23\\  
SWIFT J1009.3-4250           & 0.3-6.0 &  1.09 &  2.51$^{+0.69}_{-0.20}$ &  29.5$^{+9.60}_{-6.60}$  & 0.22 & 12.2/9 & 2.87 \\
SWIFT J1038.8-4942           & 0.5-6.8 &  0.27 &  1.11$^{+0.13}_{-0.12}$  & 0.62$^{+0.12}_{-0.10}$   & 1.45 & 109.2/108 & 1.45 \\
SWIFT J1238.9-2720           & 2.0-7.5 &  0.67  & 1.71$^{+0.70}_{-0.60}$  & 59.2$^{+25.5}_{-20.2}$   & 0.53  & 28.5/30 & 3.92 \\ 
\tableline
\end{tabular}
Notes: $\dagger$: cm$^{-2}$; $\ddagger$: erg cm$^{-2}$ s$^{-1}$, if the source has a flux variability this is a mean flux;\\
       $\blacktriangle$: source fitted with a model consisting of two power laws (only one intrinsically absorbed) having the same photon index 
       and relative normalization one 0.6\% of the other;
       an equally good fit is obtained with the unabsorbed power law component substituted by a black body model with kT $\sim$ 0.2 keV and 
       a luminosity of $\sim$ 10$^{41}$ erg s$^{-1}$.

\end{center}
\end{table}

\end{document}